\documentclass[12pt]{article}

\usepackage{graphicx}
\usepackage{epsfig}
\usepackage[usenames]{color}
\usepackage{epsfig, bm}
\usepackage{latexsym}
\usepackage{bbding}
\usepackage{amssymb}
\usepackage{ifsym}
\usepackage{wasysym}

\RequirePackage{color}
\definecolor{MyDarkGreen}{rgb}{0.02,0.60,0.06}

\title{Concentration versus dispersion of research resources: a contribution to the debate}
\author{ 
 {\it R.~Kenna$^{\,1}$} and {\it B.~Berche$^{\,2}$,} \\~\\
$^1$ Applied Mathematics Research Centre,
Coventry University,\\
Coventry, CV1 5FB, England
{}\\~\\
$^2$ Statistical Physics Group,
 Institut Jean Lamour\footnote{Laboratoire associ\'e au CNRS UMR 7198} ,\\
 CNRS -- Nancy Universit\'{e} -- UPVM, B.P. 70239,\\
 F -- 54506 Vand{\oe}uvre l\`es Nancy Cedex, France
{}\\~\\}
\begin{document}
\maketitle

{\Large
  \begin{abstract}
Using the results of the UK's research assessment exercise, we show that the size or mass
of research groups, 
rather than individual calibre or prestige of the institution, 
is the dominant factor which drives the quality of research teams. 
There are two critical masses in research: a lower one, below which teams are vulnerable and an 
upper one, above which average dependency of research quality on team size reduces.
This levelling off refutes arguments which advocate ever increasing concentration of research support
into a few large institutions.  
We also show that to increase research quality, policies which nourish two-way communication links between 
researchers are paramount.
  \end{abstract} }
%
  \thispagestyle{empty}
%
%
  \newpage
%
                  \pagenumbering{arabic}

\section{Introduction}

Research evaluation systems such as the UK's {\emph{Research Assessment Exercise (RAE)}\/} form bases
on which   governments and funding councils formulate policies on where to focus investment.
Due to some expectations that  higher quality research is generated by larger teams, there have been 
campaigns to concentrate funding on institutions which already have a wealth of resources, 
in terms of finances and staff numbers \cite{Ha09}. 
However, the most recent UK exercise, the results of which were announced 
in 2009, has identified pockets of research excellence in smaller universities as well.
This has enhanced counter-arguments by supporters of competition, who advocate a more even dispersion
 of resources  \cite{Ha09}.

The notion of {\emph{critical mass}\/} in research has existed  for a long time without 
precise definition. It is loosely described as the minimum size a research team must attain for it to be viable 
in the longer term \cite{Ha09}. 
Arguments extending this notion to larger teams have been used to support the viewpoint that bigger is better in research \cite{Ha09}.
Using the UK research base as a test bed, if a continual policy of  concentration of funding were indeed to lead to better 
quality research, one would expect that the research quality of the {\emph{Russell Group}\/} of larger universities 
with larger research teams would be superior to that of the {\emph{1994 Group}\/}, which contains smaller universities with smaller teams. 
One would expect this to be reflected, for example, in a significant difference in the average citations counts associated 
with researchers from each group. 
However, a recent report found that this was not to be the case in the main \cite{AdGu10}. 

The first  aim of this paper is to explain this recent finding. 
The explanation comes from the existence of not one, but {\emph{two}\/} critical masses, which are  discipline dependent \cite{us1}. 
The lower of these matches the heretofore loose  notion of critical mass described above, and
the research quality of teams up to about twice this size is strongly mass dependent.
However, once the quantity of researchers in a team exceeds an 
{\emph{upper critical mass}\/}, a crossover occurs and 
the dependency of quality on quantity  
reduces significantly. 
Here it is shown that the existence and properties of this second critical mass 
are the reasons why the research quality of the Russell Group and the 1994 
Group are of comparable levels.
The consequences of this is that a continual policy of concentration of resources into the largest universities is ineffective.
Instead, medium sized research groups should be strengthened to achieve upper critical mass, 
resulting in a greater collective benefit for a given discipline.

While the first aim of this paper is thus to address policy issues at the level of government and funding bodies, the
second aim concerns policy at the level of universities and teams. 
In particular, it is shown that 
two-way communication links are the main drivers of research quality, and therefore, maximisation
of the former optimises the latter. 

In this paper we describe a mathematical model introduced in Ref.~\cite{us1} and which relates the quality of 
research  to the quantity of individuals in the team.
Then, we present a brief description 
of the RAE and of how its results can be used to 
compactly reflect the quality of a research team.
The various university representation groupings in the UK are also described, and we
demonstrate how our model is capable of capturing the research quality of each such grouping simultaneously.
Our main results concern the reduced dependency of average research quality 
on team size beyond the upper
critical mass which explains 
the recent findings reported in Ref.~\cite{AdGu10}. 

\section{The Relationship between Quality and Quantity in Research}

We are interested in the relationship between the quality of research and the resources available, 
specifically in the form of the quantity of individuals in a research team.
There are two competing viewpoints in the current debate on the nature of this  relationship  \cite{Ha09}.
The first is that bigger is  always better, and support should be concentrated in a few institutes which 
already have abundant resources. 
The second viewpoint is that it is the quality of individuals that drives research.
This viewpoint is supportive of a policy of  spreading of resources to wherever excellent individuals 
are found and to support competition more evenly. 
On the other hand, according to a theory recently advanced in Ref.~\cite{us1} research quality is 
strongly quantity dependent only up to a point, beyond which this 
relationship reduces. We now summarise the basic reasoning behind these viewpoints.

Naively, one may expect that the {\emph{strength}} of a research team is approximately proportional to the
number of individuals it contains: a research group of ten individuals, say, may be expected, on average, to produce 
twice as many papers, train twice as many PhD students and generate twice as much income as a group of five.
Representing the research strength of the $i^{\rm{th}}$ individual in a research team by $\alpha_i$, the 
combined strength of a team of size $N$ according to this view is  $S = \sum_{i=1}^N{\alpha_i}$.
Defining research {\emph{quality}\/} $s=S/N$ as the average strength per team member, one has $s = \bar{\alpha}$,
where $\bar{\alpha}$ is the mean individual calibre. 
According to this naive expectation, different
teams  are thus supposed to have qualities distributed around an average
 $a$, the mean calibre for all teams in the discipline.
All particular influences, such as the prestige of the institution,
the impact of international collaborations, the presence of outstanding 
scientific personalities in the team, etc, enter the model as {\em noise}, so that 
$s=a+\hbox{noise}$.  The expected research quality in a given discipline, 
averaged over all institutions, may then be written
\begin{equation}
{\bar{s}} = a\,,
\label{intensive}
\end{equation}
which is independent of $N$. 
Borrowing terminology from physics, this naive expectation is that research quality is {\emph{intensive}\/}.
This viewpoint leads to the conclusion that the quality of research produced by a given team
is a direct measure of the strength of the individuals constituting that team and that 
individual calibre is the dominant force which drives research quality.
On this basis, the best policy to maximise the quality of a team is to recruit members
of high individual calibre to maximise the value $\bar{\alpha}$ for the team.
Here it is demonstrated that this viewpoint is too naive and a more sophisticated one is advocated.

In Ref.~\cite{us1}, an alternative, hierarchical theory for team strength was advanced. 
This has its origins in the statistical physics of {\emph{complex systems}\/}, the properties of which are not simply
the sums of the properties of their individual components. Instead {\emph{interactions}\/} between these 
components must be taken into account.
Denoting the strength of the  interaction between the  $i^{\rm{th}}$ and $j^{\rm{th}}$  
individuals in a research team as 
$\beta_{\langle{ij}\rangle}$,
the overall team strength for a sufficiently small team is now modelled by
$
 S = \sum_{i=1}^N{\alpha_i} + \sum_{{\langle{ij}\rangle}=1}^{N_{\rm{int}}}{\beta_{\langle{ij}\rangle}},
$
where, provided the team is not large (see below), the number of  two-way communication links is $N_{\rm{int}} = N(N-1)/2$.
However, there is a limit to the number of two-way communication links an individual can sustain in a large team.
If the average such limit is denoted $N_c$, then a large team of size $N>N_c$ may fragment into $n$ smaller 
subteams, which themselves may interact. In this case $N_{\rm{int}} = N N_c/2$.
With an average intra-subteam interaction strength $\bar{\beta}$, and an average inter-subteam 
interaction strength $\bar{\gamma}$, say, the average strength of a team of size $N$ is now
$
 S = \bar{\alpha}N + \bar{\beta}N_{\rm{int}} +  \bar{\gamma}n_{\rm{int}},
$
where $n_{\rm{int}} = n(n-1)/2$ is the number of inter-subteam interactions.
Therefore the expected,  relationship between research quality $s=S/N$ and 
team quantity $N$ may be modelled by \cite{us1}
\begin{equation}
\langle{s}\rangle = \left\{ \begin{array}{ll}
             a_1 + b_1 N &  {\mbox{if $N \le N_c$}} \\
             a_2 + b_2 N &  {\mbox{if $N \ge N_c$}},
             \end{array}
     \right.
\label{extensive}
\end{equation}
where $a_1$, $b_1$, $a_2$ and $b_2$ are related to the various mean interaction strengths between hierarchies.
We refer to $N_c$ as the {\emph{upper critical mass}\/}.
In fact, $b_2 \sim 1/N_c$ and is small for large $N_c$  \cite{us1}.
For this model, research  quality is, in fact, {\emph{extensive}} - it depends on the quantity of individuals 
$N$ involved in the activity. However, this dependency reduces beyond the upper critical mass,
becoming more  intensive for sufficiently large $N_c$.

The model (\ref{extensive}) allows for a definition of another critical mass  in research \cite{Ha09},
which we refer to as the {\emph{lower critical mass}\/}.
This captures the traditional notion of critical mass described in the introduction.
By considering the predicted effects of adding new members of staff to a research team, or 
of transferring staff between teams of various sizes, a scaling relation between the  lower and upper critical masses,  $N_k$ and $N_c$,
was found. This relationship is \cite{us1}
\begin{equation}
 N_k = \frac{N_c}{2}.
\label{Nk}
\end{equation}
We define  a research team of size $N$  to be 
\begin{eqnarray*}
 {\mbox{small or subcritical if}} &  &  N \le N_k, \\
 {\mbox{medium if}} &  & N_k \le N \le N_c, \\
 {\mbox{large or supercritical if}} &  & N \ge N_c. 
\end{eqnarray*}
In Ref.~\cite{us1}, it was also shown that, in order to maximise the overall strength of a research discipline, it is best to 
prioritise support for medium teams, while small teams must strive to surpass the lower critical mass to survive.
Of course, while team strength also increases with increasing calibre of individuals, this is not the
dominant mechanism. In fact, it is an order of magnitude smaller than the collaborative effect.

In Ref.~\cite{us2}, critical masses were determined for a multitude of 
research areas on the basis of the 
quality measurements coming from the UK's most recent RAE. 
Using hypothesis testing, model (\ref{intensive}) was rejected in favour of model (\ref{extensive}).
The resulting critical masses are listed in Table~1, alongside the estimates for the parameters $a_1, \dots, b_2$ for the
 disciplines analysed.
While most data sets are normal, some  fail either the Kolmogorov-Smirnov and/or the Anderson-Darling tests
and these are flagged in the table. Confidence intervals associated with fits to these data must be 
treated carefully as approximate only.

If the breakpoint were absent, the linear relationship between research quality and team quantity in the first
part of (\ref{extensive}) would be expected to extend indefinitely. In this circumstance,
maximisation of research quality would indeed be achieved by an unlimited policy of concentration.
However, as evidenced in Ref.~\cite{us2}, and as we shall see in the next section, 
evidence for the existence of the upper critical point is overwhelming.

\section{The levelling of research quality}

The UK's 2008 RAE is considered to be the most precise evaluation of its kind to date. 
This exercise was not based on citation counts. 
Instead research areas were scrutinized by experts in various fields to determine the proportion 
which fell into five quality levels. These are defined as 4* (world-leading research),
3* (internationally excellent),  2* (recognised internationally), 
1* (recognised nationally) and unclassified.
In 2009, a formula based on the resulting quality profiles
was used to determine how research funding is distributed to each university.
The formula used by the funding council for England associates each
rank with a weight in such a way that 4* and 3* research  
respectively receive seven and three times the amount of funding 
allocated to 2* research. Research ranked at or below 1* attract no funding.
This funding formula may therefore be considered as a measure of the quality $s$ of a research team.
Denoting the percentage of a team's research which was evaluated as $n*$ by $p_{n*}$, 
we define the quality of that team by $s =  p_{4*} + 3p_{3*}/7 + p_{2*}/7$.
In this way, the theoretical maximum quality is $s=100$. 
In fact no team achieved such a score, with the best teams achieving about half this.

The UK's academic research base is organised into a number of representation groups (see e.g., Ref.~\cite{Ne09} for an overview). 
These are
 (i) the {\emph{Russell Group}\/} of research intensive universities, mostly with medical schools,
(ii) the {\emph{1994 Group}\/} of research intensive universities mostly without medical schools,
(iii) the {\emph{Million+ Group}\/} of modern universities which were formed after 1992,
(iv) {\emph{University Alliance}\/} of business-like universities
(v) the {\emph{GuildHE}\/} education-focused group and the remaining 
(vi) unaffiliated universities.
As mentioned in the introduction, the result of Ref.~\cite{AdGu10} 
(which is perhaps surprising to proponents of a policy of concentration) is that,
based on citation counts, there is little difference between the research quality of the Russell
Group and the 1994 Group.

We begin the explanation of why this is the case by the sequence of plots in Fig.1, for physics.
In Fig.1(a) we normalise the quality measurements to the mean coming from Eq.(\ref{intensive})
by plotting $s-\bar{s}$ against the names of the various institutions, listed alphabetically.
For physics,  the mean measured quality of research teams  in the UK is $\bar{s}=35.9$.
From Fig.1(a), the research teams in the Russell and 1994 Groups mostly have 
quality values lying above this mean while those of the remaining universities mostly lie below. 
The nature of the situation is better revealed, however, in Fig.1(b), where the same data are plotted against the size of the
research teams. 
The solid line is a piecewise linear regression fit to the model (\ref{extensive})
and the dashed curves represent the resulting $95\%$ confidence intervals.
The correlation between quality and quantity to the left of the breakpoint is evident,
but this dependency reduces on the right.
A statistical analysis of this and other fits and the resulting $P$ values for the model are detailed in Ref.~\cite{us2}
where the coefficients of determination are also given.

The dotted line in Fig.1(b) is the extrapolation of the left fit into the supercritical region. 
In the absence of a breakpoint,
if the interactions  which govern research quality for the small and medium universities 
(described by the first part of Eq.(\ref{extensive})) applied
also to the large ones, then the research quality for the latter may also be expected to follow this line.
In this case, a policy of concentration of resources could be justified.
Clearly this is not the case.

The reason for the comparable qualities of the Russell Group and the 1994 Group is now clear from Fig.1(b). 
The large research teams in both representation groups have a different interaction pattern than 
those for small and medium groups. With a large value of $N_c=25 \pm 5$,
research quality is saturated to the right of the breakpoint, the concentration of more staff into 
these teams only leads, on average, to a linear 
increase in research strength and therefore does not significantly increase overall average research quality. 

The RAE quality results for the other representation groups are also elucidated in Fig.1(b):
they are scattered about a line of positive slope for $N < N_c$. In these cases, the addition of more 
mass, in the form of new staff, to these teams is expected to drive up quality as 
the number of two-way communication links within the team increases quadratically.
A policy of supporting medium sized groups is  expected to enhance
the quality of research in the sector overall \cite{us1}. 

The effectiveness of the model is reinforced in Fig.1(c), 
which is on the same scale as Fig.1(a) to facilitate comparison. 
In Fig.1(c)  the quality scores  have been renormalised by plotting $s-\langle{s}\rangle$ 
against the alphabetically arranged research teams, where $\langle{s}\rangle$ is the expected quality value coming from the model 
(\ref{extensive}) and is $N-$dependent. 
The standard deviations for Fig.1(a) and Fig.1(c) are $7.8$ and $5.3$, respectively, the tighter distribution about
 model (\ref{extensive})
compared to (\ref{intensive}), illustrating its superiority.
Moreover, in contrast to Fig.1(a), the data for all representation groups and for the teams belonging to unaffiliated universities straddle the line in Fig.1(c).
The model successfully captures the quality of {\emph{all}} groupings and may form the basis of a renormalised ranking system, 
which takes size into account.

Similar analyses may be performed for other research areas and those for biology,
geography, Earth and environmental sciences, archaeology, law, education, applied mathematics and sociology are given in Figs.2-8. 
(In the cases where two or more institutions put forward a joint RAE submission, that submission is associated
with the first group in the list $\{$Russell, 1994, Million+, Alliance, GuildHE, unaffiliated$\}$ to which at least 
one of them belongs.)
In each case the comparable levels of research quality 
associated with the large Russell and 1994 Groups may be explained by the existence of the upper critical point
and the levelling of the dependency of quality on quantity in the supercritical zone where  $N > N_c$.

The fitting procedure resulted in three possible values for the critical mass
in the computer sciences, and these are labelled in the table with indices 1, 2 and 3.
The coefficients of determination for these fits were $R^2=41\%$, $R^2=43\%$, and $R^2=45\%$, respectively. 
The competing  nature of these fits may be explained
if computer science is not one but several subject areas, each with their individual work patterns.
Similarly, for archaeology, the coefficient of determination for the first listed fit is $R^2=74.7\%$, and the 
second has $R^2=74.9\%$.

Also, in the second panels of each figure the line of best fit
for small and medium groups is extrapolated into the supercritical zone. 
Clearly the large groups are not described by this extrapolated line,
and this is overwhelming evidence for the existence of the upper critical mass.
However, in the case of biology (Fig.2(b)) for example, it is interesting to note that
a few of the best performing research teams, which appear as outliers to
the overall fit, are well described by this overshoot. 
These teams have sizes only marginally above $N_c$.
A similar, if less pronounced, phenomenon occurs with many of the other disciplines.
While one must be careful not to attempt to explain too much on the basis of a simple model,
and  there are undoubtedly many more complex factors at work,
it is tempting to speculate that 
this ``overshoot phenomenon'' may be caused by a greater than usual degree of cohesiveness in these 
highly successful research teams,
in which two-way communication links are sustained despite their group sizes exceeding the upper critical mass.

In the third part of each figure, renormalised plots of  $s-\langle{s}\rangle$ against $N$ are presented
for the different subject areas.
In each of these cases (and for the other subject areas listed in Table~1)
the standard deviations reduce significantly in going from the normalised plots of $s-\bar{s}$ against $N$ to
their  renormalised counterparts.
This tighter bundling of the data about the renormalised, local, quality expectation values 
indicates that the overall research base is even better than hitherto realised
as the performances of small and medium teams are closer to those of the large ones
(mostly from the Russell and 1994 Groups), once size is taken into account.

\section{Conclusions}

A current  debate within academia and between policy makers concerns the relative
merits of concentration and dispersion of research resources, and is discussed qualitatively in Ref.~\cite{Ha09}.
Here  quantitative input into this debate has been given,  which clearly supports the viewpoint that 
ever increasing concentration of resources into a small number of large institutes is
not the best way to increase overall research quality. This is because of the existence 
of an upper critical mass in research, which has been clearly established. 
Below this value, the overall strength of research teams tends to rise quadratically with increasing size,
in proportion to the number of two-way communication links. 
Beyond the upper critical mass, however, this rise reduces and approaches linearity.
Defining quality as the average strength per team member, this means that 
research quality levels out for supercritical team size.
This is the explanation behind 
recent findings based on citation counts, which show that the research 
quality from teams associated with the 1994 Group  of UK universities is on a par with 
that of the Russell Group elite \cite{AdGu10}.

The analysis presented herein also shows that simple rankings of research teams drawn up in the wake
of RAE may give a misleading impression, as they do not take size into account, and therefore
may not properly compare like with like.
Taking size into account, as in the third parts of each figure presented here,
is necessary for a better indication of team performance.

We have established that the strength of a community of interacting researchers is greater than the sum of its parts.
Having established the correlation between group size and success, and ascribing this correlation as 
primarily due to two-way communication links, it is clear that facilitation of such communication 
should form an important management policy in academia. 
For example, while modern managerial experiments such as distance working or ``hotdesking'' may be 
reasonably employed in certain industries, 
these would have a negative effect for researchers, for whom proximate location of individual office space to facilitate 
multiple, spontaneous, two-way interactions is important.
Indeed, we have seen that the best-performing research groups frequently have sizes about, or slightly 
above, the upper critical mass and we have identified a possible mechanism as to why these 
 groups outperform others.

In advance of the UK's future Research Excellence Framework, and similar exercises in other countries, 
it is  hoped that this article will help inform  debate on policy matters in the broad academic research community.

\bigskip
%

\vspace{1cm}
\noindent
{\bf{Acknowledgements}} 
Inspiring discussions with Neville Hunt, Arnaldo Donoso, Christian von Ferber and Housh Mashhoudy are gratefully acknowledged.

\newpage

\begin{table}[t!]
\caption{The results of the fit (\ref{extensive}) for a variety of academic disciplines. 
The symbol $\dagger$  indicates failure of the Kolmogorov-Smirnov normality test
and $\ddagger$ indicates that the Anderson-Darling test fails as well.
The symbol $\ast$ indicates that pure mathematics is best fitted by a single line (see Ref.~\cite{us2}).
Caveats  for the computer sciences and archaeology are discussed in the text.}
\begin{center}
\begin{tabular}{|l|r|r|r|r|r|} \hline \hline
Research discipline                                & $N_c\quad\quad$ &$a_1\quad\quad$         & $b_1\quad\quad$        & $a_2\quad\quad$      & $b_2\quad\quad$                  \\
                                         & &              &              &            &                         \\
\hline
Applied mathematics                      &$12.5 \pm 1.8$ &$ 4.8\pm 3.7$ & $2.5\pm 0.6$ & $31.7 \pm 12.8$         & $0.31\pm 0.09$    \\
Physics                                  &$25.3 \pm 4.7$ &$19.5\pm 3.3$ & $0.7\pm 0.3$ & $36.3 \pm 10.8$         & $0.05\pm 0.03$    \\
Geography/environment                 &$30.4 \pm 2.8$ &$11.8\pm 2.2$ & $1.0\pm 0.2$ & $42.2 \pm  5.6$         & $0.05\pm 0.11$    \\
Biology$^{\ddagger}$                     &$20.8 \pm 3.1$ &$-0.3\pm 5.3$ & $1.6\pm 0.4$ & $31.7 \pm 17.4$      & $0.07\pm 0.04$    \\
Chemistry                                &$36.2 \pm 12.7$&$16.3\pm 5.6$ & $0.7\pm 0.3$ & $28.4 \pm 12.7$         & $0.33\pm 0.14$    \\
Agricultural sciences$^{\ddagger}$       &$ 9.8 \pm 2.7$ &$ 2.7\pm 7.2$ & $2.3\pm 1.1$ & $23.8 \pm 23.2$      & $0.12\pm 0.06$    \\
Law$^{\dagger}$                          &$30.9 \pm 3.8$ &$ 3.3\pm 2.5$ & $1.1\pm 0.2$ & $33.0 \pm  7.3$         & $0.17\pm 0.11$    \\
Architecture/planning                 &$14.2 \pm 2.8$ &$12.9\pm 6.1$ & $1.8\pm 0.7$ & $36.1 \pm 17.5$         & $0.17\pm 0.15$    \\
French, Germanic                         &$ 6.5 \pm 0.8$ &$ 2.7\pm 6.5$ & $4.6\pm 1.4$ & $29.0 \pm 17.2$         & $0.51\pm 0.19$    \\
English                                  &$31.8 \pm 2.8$ &$ 9.6\pm 2.1$ & $1.3\pm 0.2$ & $46.8 \pm  5.3$         & $0.10\pm 0.13$    \\
Pure mathematics$^\ast$                  &$ \le 4 $      &              &              & $28.1 \pm  2.8$ & $0.5\pm 0.2$     \\
Medical sciences                         &$40.8 \pm 8.0$ &$21.7\pm 4.4$ & $0.5\pm 0.2$ & $40.8 \pm 13.2$      & $0.06\pm 0.04$    \\
Nursing, etc                             &$18.4 \pm 4.4$ &$ 8.7\pm 4.7$ & $1.1\pm 0.4$ & $26.5 \pm 13.0$      & $0.14\pm 0.16$    \\
Computer sciences 1                      &$11.3 \pm  4.7$ &$14.9\pm 7.4$ & $1.6\pm 1.1$ & $28.1 \pm 23.2$      & $0.44\pm 0.09$    \\
Computer sciences 2                      &$32.5 \pm 8.5$ &$20.3\pm 3.0$ & $0.8\pm 0.2$ & $37.3 \pm 7.7$      & $0.27\pm 0.14$    \\
Computer sciences 3                      &$49.0 \pm 10.0$ &$21.0\pm 2.6$ & $0.7\pm 0.2$ & $56.5 \pm 15.6$      & $0.01\pm 0.24$    \\
Archaelogy 1                             &$17.0 \pm 2.4$ &$13.6\pm 3.7$ & $1.7\pm 0.4$ & $39.1  \pm  10.5$      &$0.20\pm 0.14$    \\
Archaelogy 2                             &$25.4 \pm 3.2$ &$16.9\pm 3.1$ & $1.3\pm 0.2$ & $50.4 \pm  7.6$      &$-0.04\pm 0.19$    \\
Economics/econometrics                &$10.7 \pm 2.7$ &$ 0.1\pm 13.6$& $3.8\pm 1.9$ & $35.3 \pm 41.0$      & $0.49\pm 0.12$   \\
Business/management                   &$47.6 \pm 7.6$ &$11.1\pm 2.3$ & $0.6\pm 0.1$ & $33.6 \pm  6.8$      & $0.12\pm 0.06$   \\
Politics/international$^{\ddagger}$   &$25.0 \pm 4.1$ &$ 4.7\pm 4.0$ & $1.2\pm 0.3$ & $30.0 \pm 11.3$      & $0.20\pm 0.13$   \\
Sociology                                &$14.0 \pm 3.1$ &$-1.3\pm 11.3$& $2.3\pm 1.1$ & $24.6 \pm 29.6$      & $0.41\pm 0.15$    \\
Education                                &$29.0 \pm 4.4$ &$ 8.3\pm 2.2$ & $1.0\pm 0.2$ & $34.7 \pm  9.7$      & $0.05\pm 0.05$     \\
History$^{\ddagger}$                     &$24.9 \pm 4.5$ &$18.7\pm 2.6$ & $0.9\pm 0.2$ & $38.0 \pm  8.8$      & $0.16\pm 0.09$    \\
Philosophy/theology                   &$19.0 \pm 2.9$ &$13.8\pm 2.8$ & $1.7\pm 0.3$ & $43.9 \pm  8.5$      & $0.11\pm 0.19$   \\
Art \& design$^{\ddagger}$               &$25.0 \pm 7.4$ &$14.0\pm 5.1$ & $0.9\pm 0.4$ & $34.7 \pm 18.3$      & $0.04\pm 0.09$    \\
History of art, performing               &$ 8.9 \pm 1.6$ &$ 6.4\pm 7.3$ & $3.7\pm 1.1$ & $29.7 \pm 17.6$      & $1.05\pm 0.38$    \\
arts, communication                      & &              &              &             &                   \\
and music          & &              &              &             &                   \\
\hline \hline
\end{tabular}
\end{center}
\end{table}

\begin{figure*}[t]
\begin{center}
\vspace{-1cm}
\includegraphics[width=0.5\columnwidth, angle=0]{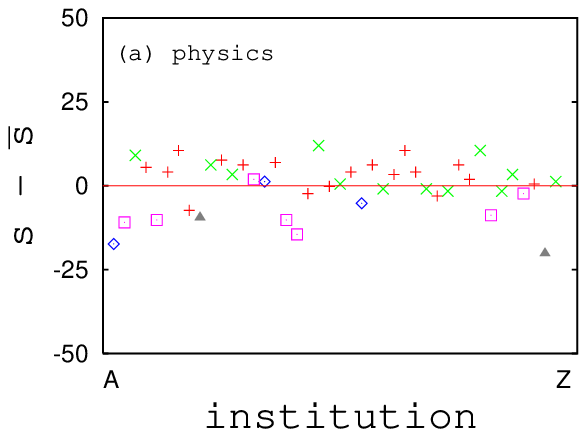}
\\
\vspace{0.6cm}
\includegraphics[width=0.5\columnwidth, angle=0]{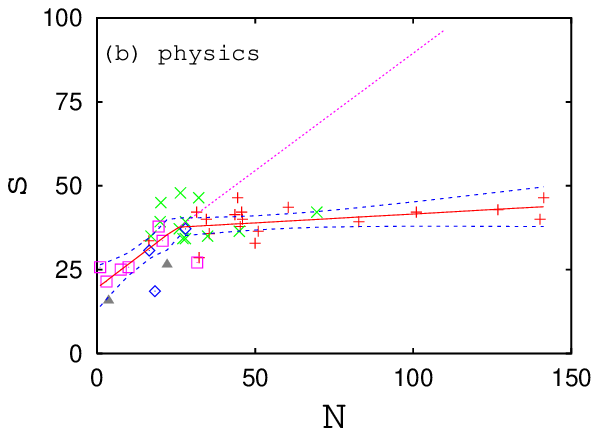}
\\
\vspace{0.6cm}
\includegraphics[width=0.5\columnwidth, angle=0]{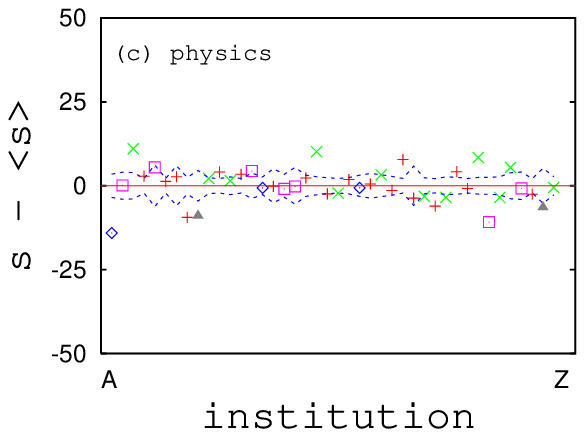}
\caption{Quality analysis of RAE results for physics. 
(a) Quality scores normalised to the overall discipline average given by (\ref{intensive})
for alphabetically listed teams.
(b) The same data plotted against the sizes $N$ of research teams where 
the solid curve is the fit coming from the model (\ref{extensive})
and the dashed curves are the corresponding 95\% confidence intervals.
The dotted line indicates how the left trend line would continue if the breakpoint
(upper critical point) were absent.
(c) Quality renormalised to the expectation values  $\langle{ s }\rangle$ coming from the model (\ref{extensive}).
The various symbols  indicate members of the
Russell Group~({\color{red}{+}}),
the 1994 Group~({\color{green}{$\times$}}),
the Million+ Group~({\color{Gray}{$\blacktriangle$}}),
the University Alliance~({\color{blue}{$\Diamond$}}),
and unaffiliated universities~({\color{magenta}{$\Box$}}).
}
\end{center}
\label{uoa19}
\end{figure*}


\begin{figure*}[t]
\begin{center}
\includegraphics[width=0.5\columnwidth, angle=0]{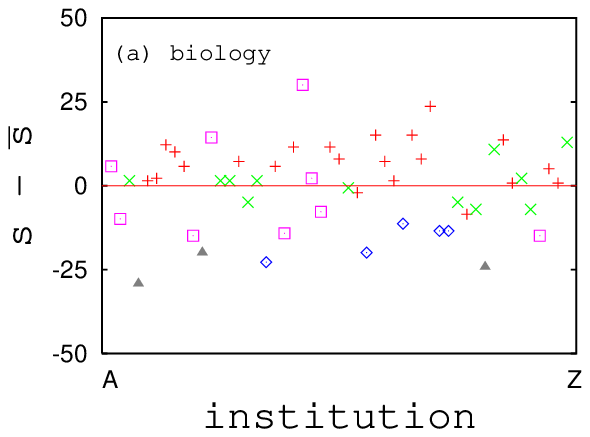}
\\
\vspace{0.6cm}
\includegraphics[width=0.5\columnwidth, angle=0]{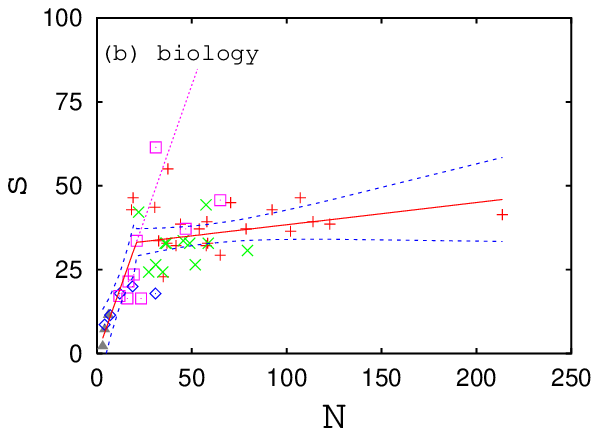}
\\
\vspace{0.6cm}
\includegraphics[width=0.5\columnwidth, angle=0]{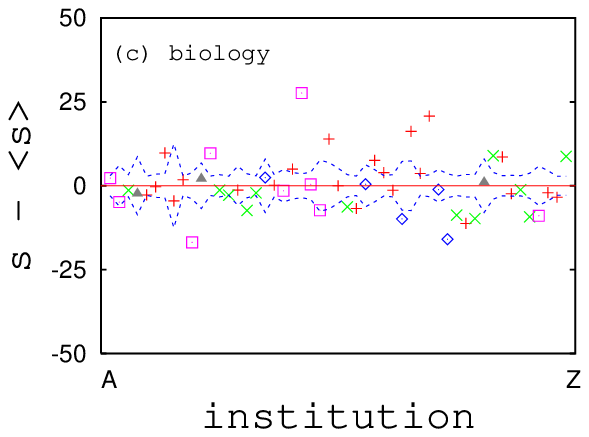}
\caption{Quality analysis for biology analogous to Fig.1. 
The standard deviation reduces from $12.5$ in panel (a) to $8.5$ in panel (c).
Note that in panel (b) the two points with maximum $s$ values, which appear as outliers to
the overall fit, are in line with the extended left fit, possibly hinting at an explanation
behind the corresponding teams' success, as discussed in the text. 
}
\end{center}
\label{uoa14}
\end{figure*}

\begin{figure*}[t]
\begin{center}
\includegraphics[width=0.5\columnwidth, angle=0]{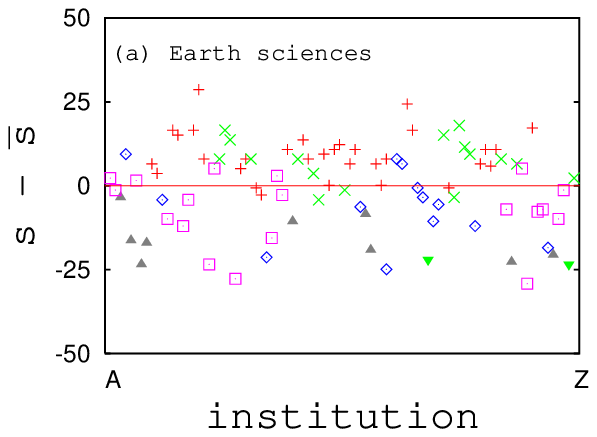}
\\
\vspace{0.6cm}
\includegraphics[width=0.5\columnwidth, angle=0]{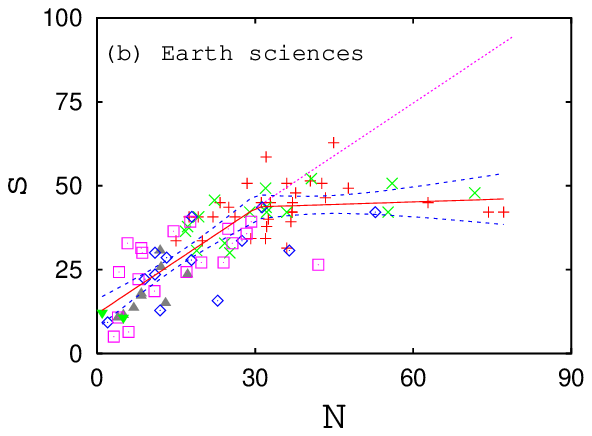}
\\
\vspace{0.6cm}
\includegraphics[width=0.5\columnwidth, angle=0]{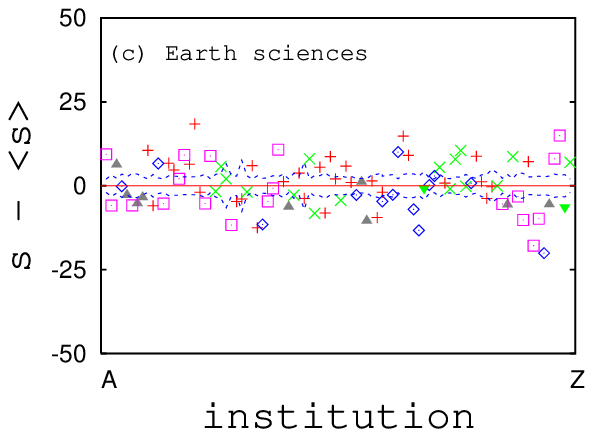}
\caption{Quality analysis for geography, Earth and environmental sciences analogous to Fig.1. 
The standard deviation reduces from $12.8$ in panel (a) to $7.5$ in panel (c).
The various symbols (in colour online) indicate members of the
Russell Group~({\color{red}{+}}),
the 1994 Group~({\color{green}{$\times$}}),
the Million+ Group~({\color{Gray}{$\blacktriangle$}}),
the University Alliance~({\color{blue}{$\Diamond$}}),
Guild HE~({\color{green}{$\blacktriangledown$}}),
and unaffiliated universities~({\color{magenta}{$\Box$}}).}
\end{center}
\label{uoa3217}
\end{figure*}

\begin{figure*}[t]
\begin{center}
\includegraphics[width=0.5\columnwidth, angle=0]{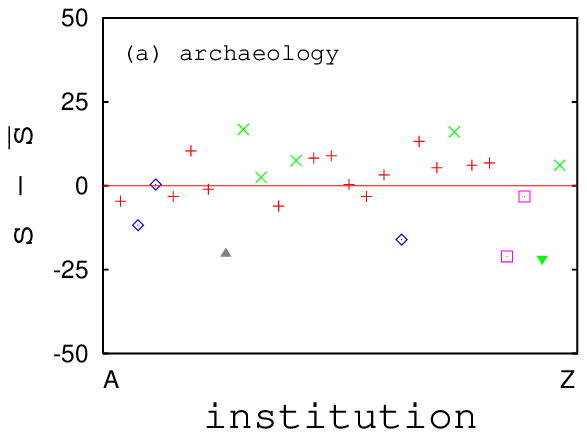}
\\
\vspace{0.6cm}
\includegraphics[width=0.5\columnwidth, angle=0]{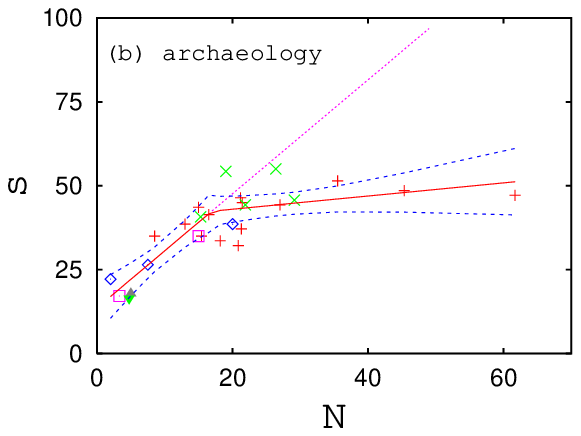}
\\
\vspace{0.6cm}
\includegraphics[width=0.5\columnwidth, angle=0]{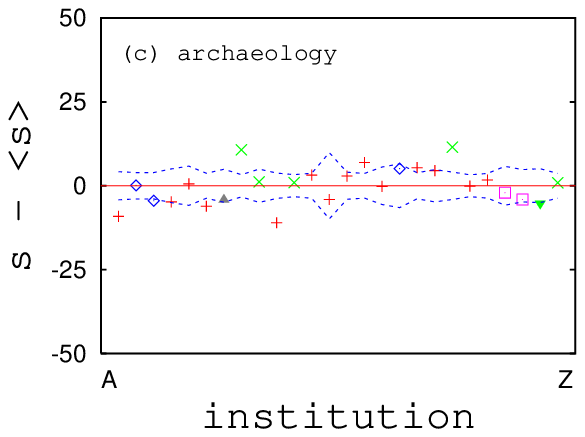}
\caption{Quality analysis for archaeology with breakpoint at $N_c=17.0\pm 2.4$. 
The standard deviation reduces from $11.0$ in panel (a) to $5.5$ in panel (c).
The figure for the larger breakpoint value listed in the table is similar, with the same 
reduction in standard deviations.}
\end{center}
\label{uoa33reduced}
\end{figure*}

\begin{figure*}[t]
\begin{center}
\includegraphics[width=0.5\columnwidth, angle=0]{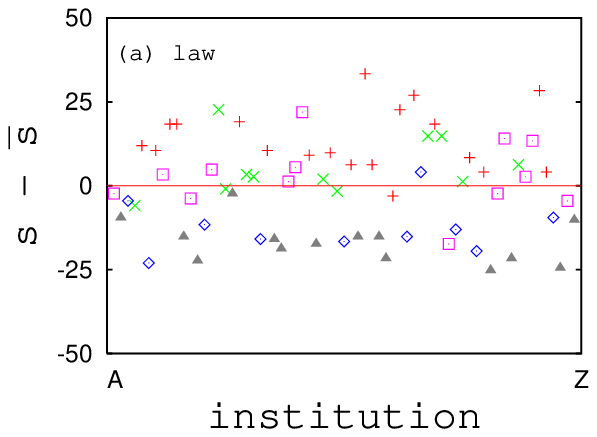}
\\
\vspace{0.6cm}
\includegraphics[width=0.5\columnwidth, angle=0]{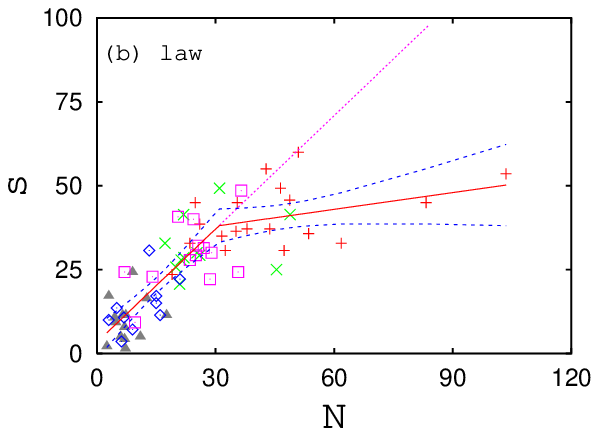}
\\
\vspace{0.6cm}
\includegraphics[width=0.5\columnwidth, angle=0]{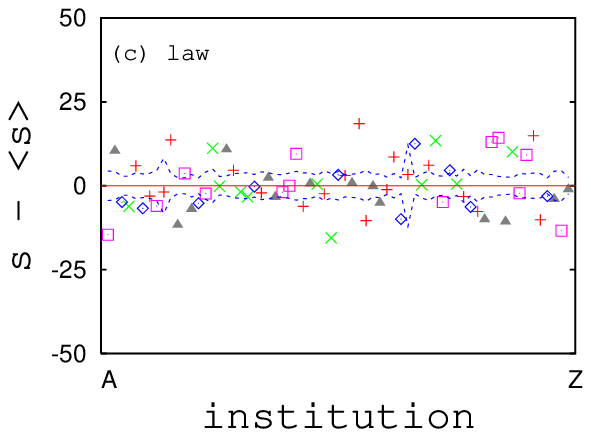}
\caption{Quality analysis for law analogous to Fig.1. 
The standard deviation reduces from $14.7$ in panel (a) to $7.9$ in panel (c).
}
\end{center}
\label{uoa38}
\end{figure*}

\begin{figure*}[t]
\begin{center}
\includegraphics[width=0.5\columnwidth, angle=0]{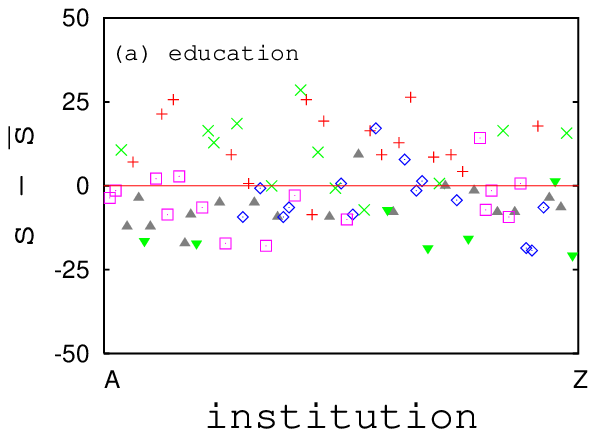}
\\
\vspace{0.6cm}
\includegraphics[width=0.5\columnwidth, angle=0]{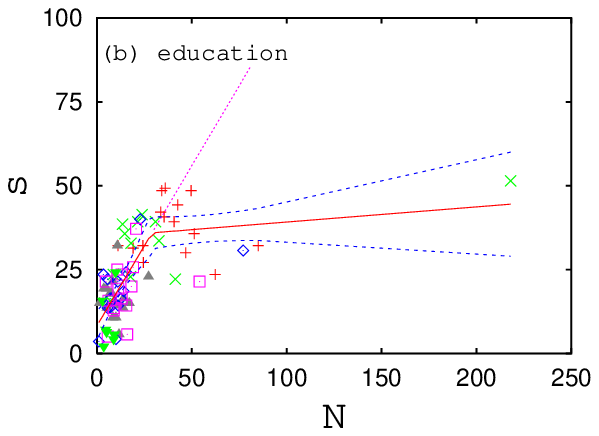}
\\
\vspace{0.6cm}
\includegraphics[width=0.5\columnwidth, angle=0]{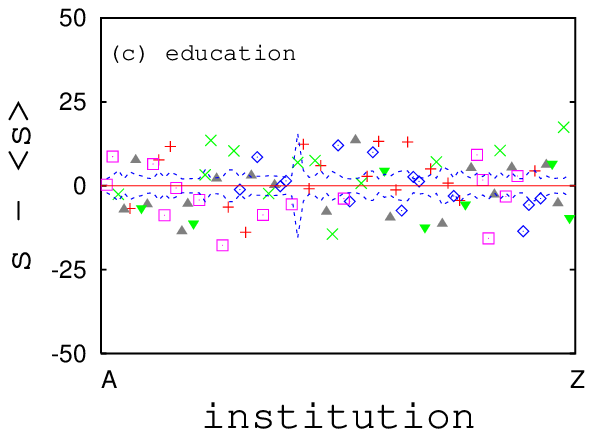}
\caption{Quality analysis for education analogous to Fig.1. 
The standard deviation reduces from $12.2$ in panel (a) to $8.1$ in panel (c).
}
\end{center}
\label{uoa45}
\end{figure*}

\begin{figure*}[t]
\begin{center}
\includegraphics[width=0.5\columnwidth, angle=0]{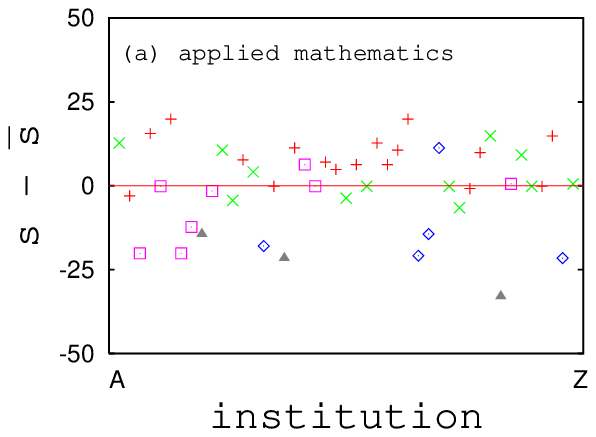}
\\
\vspace{0.6cm}
\includegraphics[width=0.5\columnwidth, angle=0]{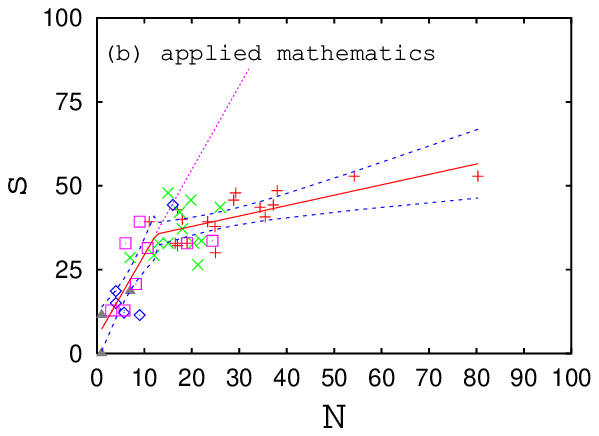}
\\
\vspace{0.6cm}
\includegraphics[width=0.5\columnwidth, angle=0]{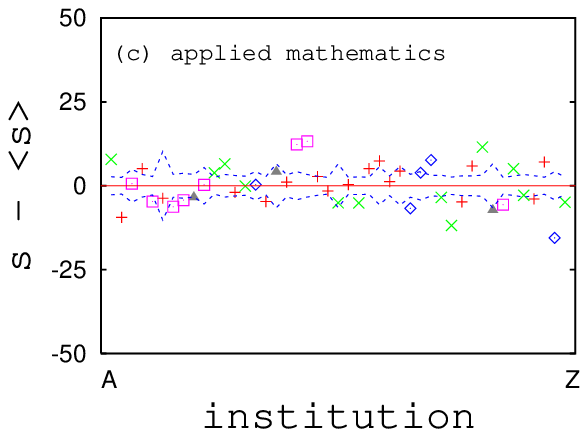}
\caption{Research quality analysis for applied mathematics.
The standard deviation reduces from $12.6$ about model (\ref{intensive}) and corresponding to panel (a) 
to $6.4$  about model (\ref{extensive}), corresponding to panel (c).
}
\end{center}
\label{uoa21}
\end{figure*}

\begin{figure*}[t]
\begin{center}
\includegraphics[width=0.5\columnwidth, angle=0]{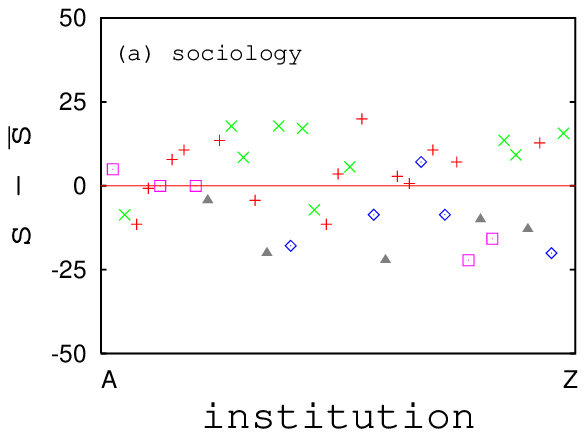}
\\
\vspace{0.6cm}
\includegraphics[width=0.5\columnwidth, angle=0]{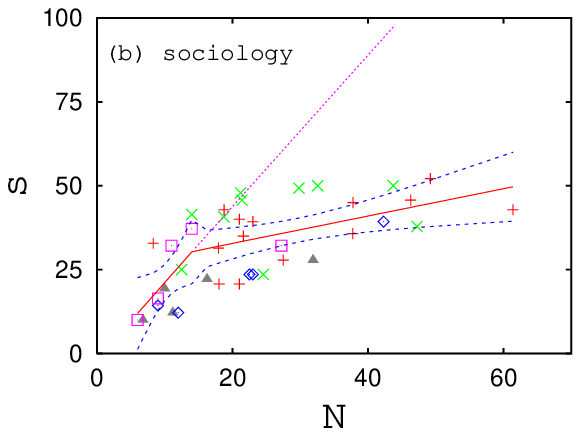}
\\
\vspace{0.6cm}
\includegraphics[width=0.5\columnwidth, angle=0]{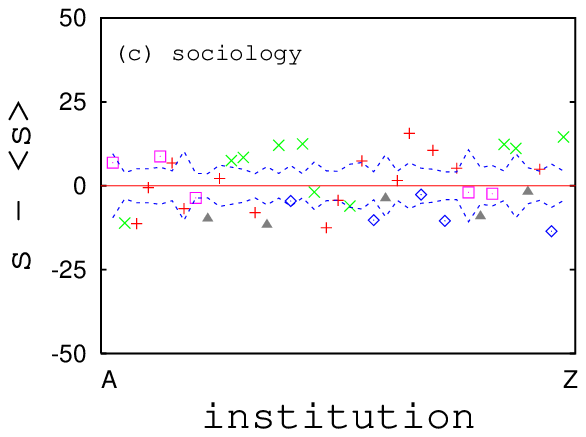}
\caption{Quality analysis for sociology analogous to Fig.1. 
The standard deviation reduces from $12.5$ in panel (a) to $8.8$ in panel (c).
}
\end{center}
\label{uoa41}
\end{figure*}

\end{document}